\documentclass[preprint,proceedings]{rmaa}

\usepackage{paralist}
\usepackage{psfrag,color}

\SetYear{2002}
\SetConfTitle{Galaxy Evolution: Theory and Observations}

\title{New Problems for the Formation of Disk Galaxies} 

\author{Frank C.~van den Bosch \affil{Max-Planck Institute for
Astrophysics, Garching, Germany}}

\shortauthor{van den Bosch}
\shorttitle{New Problems for the Formation of Disk Galaxies}

\suppressfulladdresses
\fulladdresses{
\item Frank C. van den Bosch: Max-Planck Institute for Astrophysics,
  Karl-Schwarzschild Str. 1, 85741 Garching, Germany
 (vdbosch@mpa-garching.mpg.de).}

\listofauthors{Frank C.~van den Bosch}
\indexauthor{van den Bosch, F. C.}

\abstract{I discuss the  role of angular momentum in  the formation of
disk  galaxies, and  describe  the  results of  two  studies aimed  at
testing the standard paradigm for disk formation.}

\addkeyword{dark      matter}     \addkeyword{galaxies:     formation}
\addkeyword{galaxies: structure}

\begin{document}

\maketitle

\section{Introduction}
\label{sec:intro}

The  current  paradigm for  disk  formation  contains three  important
ingredients:  (i) the  angular momentum  originates  from cosmological
torques (ii)  the gas and  dark matter within virialized  systems have
initial angular  momentum distributions (AMDs) that  are identical and
(iii) the  gas conserves its  specific angular momentum  when cooling.
Under these  assumptions the predicted scale lengths  of disk galaxies
are in excellent agreement with observations (Fall \& Efstathiou 1980;
de Jong \&  Lacey 2000), which has motivated  the construction of more
detailed models,  but always under the three  assumptions listed above
(e.g., Mo,  Mao \& White 1998;  van den Bosch 1998,  2000, 2001, 2002;
Firmani \& Avila-Reese 2000).

Because of  the overall success of  these models in  explaining a wide
range of observed  properties of disk galaxies, it  has generally been
assumed  that the  aforementioned assumptions  are  correct.  However,
several  recent results  have started  to cast  some doubt  as  to the
validity  of   this  standard  framework.   First   of  all,  detailed
hydro-dynamical simulations  of disk formation  in a cold  dark matter
(CDM) Universe  yield disks that are  an order of  magnitude too small
(e.g., Steinmetz \& Navarro 1999).  This problem, known as the angular
momentum catastrophe,  is a consequence of  the hierarchical formation
of galaxies which causes the baryons to lose a large fraction of their
angular momentum to the dark matter.

Secondly, under assumption (iii)  the density distribution of disks is
a direct  reflection of  the AMD in  the proto-galaxy.   Bullock \etal
(2001, hereafter  B01) determined the  AMDs of individual  dark matter
halos, which according to assumption  (ii) should be identical to that
of  the  gas,   and  thus  to  that  of   the  disk.   However,  these
distributions seem to have far  too much low angular momentum material
for consistency with the  typical exponential density distributions of
disk galaxies (B01; van den Bosch 2001).
\begin{figure}[!t]
  \includegraphics[width=\columnwidth]{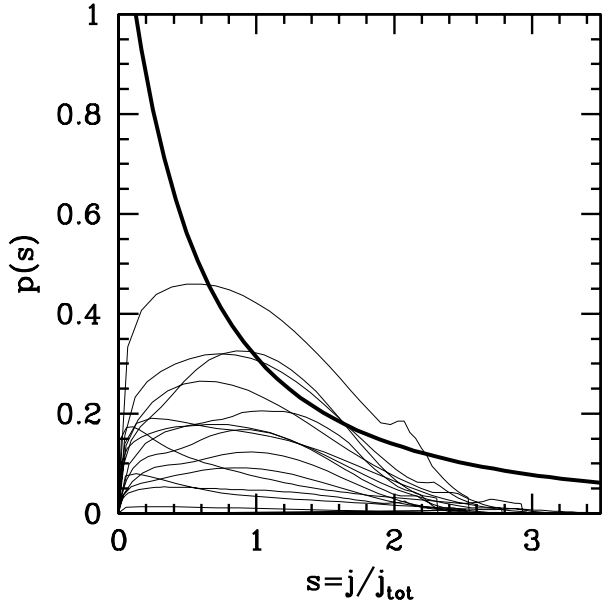} 
  \caption{A comparison of  the AMDs of 14 disk  galaxies (thin lines)
  with  those  of CDM  halos  (thick  line)  as parameterized  by  the
  `universal'  profile  introduced  by  B01.  Whereas  the  latter  is
  normalized to unity, the former  are normalized to the ratio of disk
  mass to  expected baryon mass  (i.e., the universal  baryon fraction
  times the total  virial mass). As is evident,  only a small fraction
  of the baryons within the halo's  virial radius have ended up in the
  disk,  and  with  a AMD  that  strongly  differs  from that  of  the
  `universal' distribution for CDM halos.}  \label{fig:amd}
\end{figure}
%


\section{Testing the paradigm}

If assumption (ii) and (iii)  are correct, {\it and all baryons inside
the virial  radius make it  into the disk},  the AMD of a  disk galaxy
should be identical to that  of the ``universal'' profile suggested by
B01.  In  van den  Bosch, Burkert \&  Swaters (2001), we  computed the
AMDs of 14 dwarf galaxies for which accurate photometry and kinematics
are   available.   The   results   are  shown   in   Figure  1.    Two
characteristics  are apparent.  First  of all,  only a  small fraction
(between 2 and  75 percent) of the available baryons  have ended up in
the disk:  the remaining gas  has either not  yet cooled, or  has been
expelled from the disk  through feedback processes. Secondly, the AMDs
of the disk differ strongly from  that of the dark matter halos.  This
suggests that  either preferentially the low-angular  momentum gas has
been prevented from  ending up in the disk galaxy or  the gas and dark
matter started out with different angular momentum distributions.  The
former might be established by  a specific kind of feedback, while the
latter would imply inconsistency  with assumption (ii) of our standard
framework for disk formation.

Therefore, in  order to  test whether indeed  the gas and  dark matter
acquire identical AMDs van  den Bosch \etal (2002) performed numerical
simulations  of  structure   formation  in  a  $\Lambda$CDM  cosmology
including both  dark matter  (DM) and a  non-radiative gas.   For each
halo in this  simulation we computed the AMDs of both  the gas and the
associated dark  matter. A detailed investigation of  these AMDs leads
to the following two conclusions:
\begin{itemize}
\item The gas and DM have virtually identical AMDs 
\item Between 10 and 40 percent of the gas has negative specific
angular momentum (w.r.t. the total angular momentum vector).
\end{itemize}
The former indicates that assumption  (ii) in the standard paradigm of
disk formation is correct. However,  since disk galaxies in general do
not  contain significant  amounts of  counter-rotating  material, {\it
disk formation  cannot occur  under detailed conservation  of specific
angular momentum of the baryons}, in conflict with assumption (iii).

\section{Discussion}

Since the seminal paper by Fall \& Efstathiou (1980), a standard model
for the formation of disk galaxies has been around that describes disk
formation  in  terms  of   the  way  gas  acquires,  and  subsequently
conserves, specific angular momentum. Surprisingly enough, very little
attention  has been  paid to  testing the  validity of  the underlying
assumptions. Via  numerical simulations we  have shown, for  the first
time, that in accord with this standard framework, gas and dark matter
acquire  identical AMDs.   However, the  fact that  these  AMDs reveal
large  mass fractions  with {\it  negative} specific  angular momentum
implies  that the  gas cannot  conserve its  detailed  distribution of
specific  angular  momentum  when  cooling  to form  the  disk.   This
crumples  one  of  the  main  pillars of  our  standard  picture,  and
indicates that a new spin  on the angular momentum acquisition of disk
galaxies may be required.

Additional puzzles come  from the fact that the  AMDs of observed disk
galaxies are  dramatically different than those of  dark matter halos:
disk predominantly  lack low angular  momentum material.  Furthermore,
detailed numerical simulations of  disk formation that include cooling
find a  large transfer of  angular momentum from  the gas to  the dark
matter, resulting in  disks that are an order  of magnitude too small.
It is  currently unclear whether these puzzles  indicate a fundamental
problem  for the  theory or  merely for  the particular  way  in which
feedback  processes  are  implemented  (or  ignored)  in  the  current
simulations and/or  models for  disk galaxy formation.   A first-order
attempt to  address the  impact of feedback  processes on  the angular
momentum  of the  gas in  proto-galaxies  is presented  in Abel  \etal
(2002),  where  it is  shown  that  any  feedback process  capable  of
expelling baryons  from dark matter  halos, is likely to  decouple the
angular momentum of the gas from that of the baryons.



\begin{thebibliography}


\bibitem{} Abel, T., van den  Bosch, F.C., \& Hernquist, L. 2002, in
preparation

\bibitem{}
Bullock, J. S., et al. 2001, \apj , 555, 240 (B01)

\bibitem{}
de Jong, R.S., \& Lacey, C. 2000, \apj , 545, 781

\bibitem{}
Fall, S. M., \& Efstathiou, G. 1980, \mnras, 193, 189 

\bibitem{}
Firmani, C., \& Avila-Reese, V. 2000, \mnras , 315, 457

\bibitem{}
Mo, H. J., Mao, S., \& White, S. D. M. 1998, \mnras , 295, 319

\bibitem{}
Sommer-Larsen, J., Gelato, S., \& Vedel, H. 1999, \apj , 519, 501

\bibitem{}
Steinmetz, M., \& Navarro, J. F. 1999, \apj , 513, 555

\bibitem{} 
van den Bosch, F. C. 1998, \apj , 507, 601

\bibitem{}
van den Bosch, F. C. 2000, \apj , 530, 177

\bibitem{}
van den Bosch, F. C. 2001, \mnras , 327, 1334

\bibitem{}
van den Bosch, F. C. 2002, \mnras , 332, 456

\bibitem{}
van den Bosch, F. C., Burkert, A., \& Swaters, R. A. 2001, \mnras ,
326, 1205

\bibitem{}
van den Bosch, F. C., Abel, T., Croft, R. A. C., Hernquist, L., \&
White, S. D. M. 2002, \apj , 576, 21 

\end{thebibliography}
\end{document}